\documentclass[traditabstract]{aa}

\usepackage[colorlinks=true, linkcolor=blue, citecolor=blue, urlcolor=blue]{hyperref}
\usepackage{amsmath}
\usepackage{graphicx}
\usepackage{txfonts}
\usepackage{amssymb}
\usepackage[]{natbib}
\usepackage{mathrsfs}
\usepackage{caption}
\usepackage{float}
\usepackage{afterpage}
\usepackage{amstext}

\newcommand{\hp}[0]{\mbox{HAT-P-26}}
\newcommand{\hpb}[0]{\mbox{HAT-P-26b}}

\begin{document}

   \title{Indications for transit timing variations in the exo-Neptune HAT-P-26b}
   \subtitle{}
   \author{C. von Essen$^{1,2}$, S. Wedemeyer$^{3}$, M. S. Sosa$^{4,5}$, M. Hjorth$^1$, V. Parkash$^8$, J. Freudenthal$^{7}$, M. Mallonn$^{9}$, R. G. Micul\'an$^{4,5}$, \mbox{L. Zibecchi$^{4,5}$}, S. Cellone$^{4,6}$, A. F. Torres$^{4,5}$} \authorrunning{C. von Essen et al. (2019)}
   \titlerunning{TTVs in the exo-Neptune HAT-P-26b}
   \offprints{cessen@phys.au.dk}
   \institute{$^1$Stellar Astrophysics Centre, Department of Physics and Astronomy, Aarhus University, Ny Munkegade 120, DK-8000 Aarhus C, Denmark\\ 
     $^2$Astronomical Observatory, Institute of Theoretical Physics and Astronomy, Vilnius University, Sauletekio av. 3, 10257, Vilnius, Lithuania\\
     $^3$Institute of Theoretical Astrophysics, University of Oslo, Postboks 1029 Blindern, N-0315 Oslo, Norway\\
     $^4$Facultad de Ciencias Astron\'omicas y Geof\'{\i}sicas, Universidad Nacional de La Plata, Paseo del Bosque, B1900FWA, La Plata, Argentina\\
     $^5$Instituto de Astrof\'{\i}sica de La Plata (CCT-La Plata, CONICET-UNLP), Paseo del Bosque, B1900FWA, La Plata, Argentina\\
     $^6$Consejo Nacional de Investigaciones Cient\'{\i}ficas y T\'ecnicas, Godoy Cruz 2290, C1425FQB, Ciudad Aut\'onoma de Buenos Aires, Argentina\\
     $^7$Institut f\"ur Astrophysik, Georg-August-Universit\"at G\"ottingen, Friedrich-Hund-Platz 1, 37077 G\"ottingen, Germany\\
     $^8$Monash Centre for Astrophysics, School of Physics and Astronomy, Monash University, Victoria 3800, Australia\\
     $^9$Leibniz-Institut f\"ur Astrophysik Potsdam, An der Sternwarte 16, 14482, Potsdam, Germany\\
     \email{cessen@phys.au.dk}
   }

   \date{Received XXXX; accepted XXXX}

\abstract{From its discovery, the low density transiting Neptune
  \hpb\ showed a 2.1$\sigma$ detection drift in its spectroscopic
  data, while photometric data showed a weak curvature in the timing
  residuals that required further follow-up observations to be
  confirmed. To investigate this suspected variability, we observed 11
  primary transits of \hpb\ between March, 2015 and July, 2018. For
  this, we used the 2.15 meter Jorge Sahade Telescope placed in San
  Juan, Argentina, and the 1.2 meter STELLA and the 2.5 meter Nordic
  Optical Telescope, both located in the Canary Islands, Spain. To add
  upon valuable information on the transmission spectrum of \hpb, we
  focused our observations in the $R$-band only. To contrast the
  observed timing variability with possible stellar activity, we
  carried out a photometric follow-up of the host star along three
  years. We carried out a global fit to the data and determined the
  individual mid-transit times focusing specifically on the light
  curves that showed complete transit coverage. Using bibliographic
  data corresponding to both ground and space-based facilities, plus
  our new characterized mid-transit times derived from
  parts-per-thousand precise photometry, we observed indications of
  transit timing variations in the system, with an amplitude of
  $\sim$4 minutes and a periodicity of $\sim$270 epochs. The
  photometric and spectroscopic follow-up observations of this system
  will be continued in order to rule out any aliasing effects caused
  by poor sampling and the long-term periodicity.}

          
   \maketitle

\section{Introduction}

From the first exoplanets ever detected around other stars than our
Sun, the most successful exoplanet detection methods have been the
radial velocity \citep[RV, such as][]{Butler2006} and the transit
\citep[][and onward]{Charbonneau2000} techniques. The main engine of
this work, the {\it Transit Timing Variation} method (TTV), gained
special relevance with the advent of the Kepler space telescope
\citep{Borucki2010,Koch2010}. For this technique, once an exoplanet is
detected via the transit method, the variations of the observed
mid-transit times allow the subsequent detection and/or
characterization of further transiting \citep{Carter2012} and
non-transiting \citep{Barros2014} exoplanets. The technique is
sensible to planets with masses as low as an Earth mass
\citep{Agol2005}, which would be extremely challenging to detect or
characterize by means of other techniques. The TTV method requires
sufficiently long baseline, precise photometry, and good phase
coverage. All these requirements are satisfied by Kepler data
\citep[see e.g.,][for a TTV characterization of hundreds of Kepler
  Objects of Interest, KOIs]{Mazeh2013}. As a consequence, surveys
focused on TTVs from the ground have been mainly carried out
photometrically following-up hot Jupiters with deep transits orbiting
bright stars
\citep[e.g.,][]{Maciejewski2011,Fukui2011,vonEssen2013}. Nonetheless,
none of them revealed unquestionable detection of TTVs. On the
contrary, TTVs in the Kepler era revealed that multiple systems are
not rare at all, and that most of the planets in these multiple
systems are within the Super Earth/mini Neptune regime
\citep[see][]{Holman2010,Lissauer2011a,Cochran2011}. Thus, from Kepler
results we can re-focus our observing capabilities and boost our
success rate from the ground by programming photometric follow-ups of
more suitable transiting systems, including the Neptune planets
instead of hot Jupiter ones.

Between 2015 and 2018 our group carried out a photometric follow-up of
three Neptune-sized exoplanetary systems. The observations were mainly
focused around primary transits. These systems presented either
``hints'' of multiplicity, or showed orbital and physical parameters
similar to KOIs in multiple systems with large amplitude TTVs. Here we
present our efforts in the detection of TTVs in \hpb, which is one of
such aforementioned systems. \hpb, a low-density Neptune-mass planet
transiting a K-type star, was discovered in 2011 by the HATNet
consortia \citep{Hartman2011}. Once the spectral reconnaissance was
carried out, additional high-resolution spectra were acquired to
better characterize the RV variations and the stellar parameters. The
spectroscopic analysis revealed a chromospherically quiet star, along
with a Neptune-mass planet orbiting its host star each $\sim$4.23
days. A combined analysis between photometric and spectroscopic data
provided better constraints on the planetary mass and radius. Further
analysis of spectroscopic data revealed a 2.1$\sigma$ detection drift
in its RVs. Nonetheless, the data were not sufficient to characterize
the system's multiplicity. Furthermore, \cite{Stevenson2016} observed
a weak curvature in the timing residuals of \hpb, but yet again
without proper confirmation due to insufficient data. Although the
\mbox{$\Delta$F = 0.53\%} transit depth makes the transit challenging
to be observed using ground-based facilities, the observed RV drift
and curvature in the timing residuals makes \hpb\ an interesting
system for TTV studies. To confront the TTV detection with other
possible sources of variability, we also carried out a photometric
follow-up of the host star along three years.

Besides TTV analysis, several efforts were produced to characterize
the chemical constituents of the atmosphere of \hpb. While
\cite{Stevenson2016} found tentative evidence for water and a lack of
potassium in transmission, \cite{Wakeford2017} measured \hpb's
atmospheric heavy element content and characterized its atmosphere as
primordial. The observations presented here were focused solely in the
$R$-band. Thus, besides the TTV discovery and characterization, in
this work we also provide an accurate wavelength-dependent
planet-to-star radii ratio around the Johnson-Cousins $R$ band,
\mbox{635 $\pm$ 100 nm}, in order to contribute with its
exo-atmospheric characterization.

In this work, Section~\ref{Sec:ODR} details our observations and
specifies the data reduction techniques. Section~\ref{Sec:Analysis}
shows the steps involved in the transit analysis and the derived model
parameters. Section~\ref{Sec:TSQ} shows our results on TTVs, and we
finish with a brief discussion and our conclusions in
Section~\ref{Sec:DaC}.

\begin{table*}[ht!]
  \caption{\label{tab:obspars} Relevant parameters of our
    observations. From left to right: the date corresponding to the
    beginning of the local night in yyyy.mm.dd, the telescope
    performing the observations, the standard deviation of the
    residual light curves in parts-per-thousand (ppt), $\sigma_{res}$,
    the number of data points per light curve, N, the average cadence
    in seconds, CAD, the total observing time, T$_{tot}$, in hours,
    the airmass range, $\chi_{max,min}$, showing maximum and minimum
    values, the $\beta$ factor, to quantify correlated noise, the
    maximum variability in pixel position, $\Delta$pix, considering
    both x and y shifts, and the maximum variability in seeing,
    $\Delta$s, in pixels. Both values are rounded up. Last column
    shows the transit coverage, TC; a description of the transit
    coding is detailed in the footnote of this table.}
  \centering
  \begin{tabular}{l c c c c c c c c c c}
    \hline \hline
    Date       & Telescope   &   $\sigma_{res}$  &    N    &   CAD   &   T$_{tot}$   &   $\chi_{max,min}$   &  $\beta$      & $\Delta$pix  &   $\Delta$s    &    TC    \\
    yyyy.mm.dd &             &   (ppt)          &         &  (sec)  &   (hours)       &                 &               &  (max$_{x,y}$) &   (pix)        &          \\
    \hline
    2015.03.30 & CASLEO      &   1.8            &  597    &  25     &  4.31        &  1.53,1.23         & 2.33          &  40            & 3              & OIBEO    \\
    2015.04.12 & NOT         &   0.8            &  843    &  18     &  4.24        &  2.33,1.10         & 2.00          &  4             & 3              & OIBEO    \\
    2015.04.16 & CASLEO      &   1.6            &  307    &  74     &  6.28        &  1.94,1.23         & 2.15          &  50            & 6              & OIBEO    \\
    2015.05.20 & NOT         &   1.8            &  482    &  19     &  2.67        &  1.82,1.13         & 1.62          &  3             & 2              & OIBE-    \\
    2015.06.06 & NOT         &   3.1            &  943    &  15     &  3.91        &  2.95,1.12         & 1.19          &  4             & 4              & OIBEO    \\
    2015.06.23 & NOT         &   2.0            &  977    &  14     &  3.81        &  2.33,1.10         & 1.47          &  2             & 2              & OIBEO    \\ 
    2016.05.14 & CASLEO      &   1.9            &  372    &  36     &  3.77        &  2.13,1.23         & 1.51          &  120           & 8              & -IBEO    \\
    2017.05.13 & CASLEO      &   3.3            &  289    &  63     &  5.12        &  2.03,1.23         & 2.59          &  52            & 3              & OIBEO    \\
    2017.05.30 & CASLEO      &   1.8            &  742    &  22     &  4.58        &  1.94,1.23         & 2.84          &  35            & 3              & OIBEO    \\ 
    2017.06.16 & CASLEO      &   1.9            &  174    &  52     &  2.53        &  1.31,1.23         & 1.42          &  30            & 3              & OIB- -   \\
    2018.07.01 & STELLA      &   1.2            &  141    & 112     &  4.42        &  1.98,1.09         & 1.75          &  350           & 6              & --BEO    \\ 
    \hline
  \end{tabular}
  \tablefoot{The letter code to specify the transit coverage during
    each observation is the following: O: out of transit, before
    ingress. I: ingress. B: flat bottom. E: egress. O: out of transit,
    after egress.}
\end{table*}

\section{Observations and data reduction}
\label{Sec:ODR}

\subsection{Observing sites and specifications of the collected data for transit photometry}

Between March, 2015 and July, 2018 we observed \hp\ \citep[\mbox{ R =
    11.56, V = 11.76,}][]{Hog2000} during ten primary transits. Our
observations were performed using the 2.15 meter Jorge Sahade
Telescope located at the Complejo Astron\'omico El
Leoncito\footnote{Visiting Astronomer, Complejo Astron\'omico El
  Leoncito operated under agreement between the Consejo Nacional de
  Investigaciones Cient\'ificas y T\'ecnicas de la Rep\'ublica
  Argentina and the National Universities of La Plata, C\'ordoba and
  San Juan; www.casleo.gov.ar} (CASLEO) in San Juan, Argentina, the
2.5 meter Nordic Optical Telescope\footnote{not.iac.es} (NOT) located
in La Palma, Spain, and the 1.2 meter STELLA, located in Tenerife,
Spain. The most relevant parameters of our observations are summarized
in Table~\ref{tab:obspars}.  All the observations were carried out
using a Johnson-Cousins $R$ filter, to minimize the impact of our
Earth's atmosphere in the overall signal-to-noise ratio of our
measurements circumventing telluric lines and the strong absorption of
stellar light caused by Rayleigh scattering. Also, several transits
allowed us to accurately characterize the planetary size within that
wavelength range. To increase the photometric precision of our data
all telescopes were slightly defocused
\citep{Kjeldsen1992,Southworth2009b}. In consequence, values of seeing
taken from science frames are not representative of the quality of the
sky at both sites. Table~\ref{tab:obspars} shows then only intra night
variability of the seeing. Exposure times were typically set between
10 and 60 seconds, and the photometric precision ranged between 1.1
and 3.7 parts-per-thousand (ppt), in all cases below the transit depth
($\sim$5.3 ppt). Seven light curves show complete transit
coverage. The rest show partial transit coverage mainly due to poor
weather conditions.

During each observing night we obtained a set of 15 bias and between
10 to 15 twilight sky flats. Due to low exposure times and optimum
cooling of charge-coupled devices, we did not take dark frames. In the
case of CASLEO data we observed using a focal reducer. This produces
an unvignetted field of view of \mbox{9 arcmin} of diameter, allowing
the simultaneous observation of \hp\ along with two reference stars of
similar brightness, namely \mbox{TYC 320-426-1} \citep[V =
  11.08,][]{Hog2000} and \mbox{TYC 320-1376-1} \citep[V =
  12.5,][]{Hog2000}, also observed by STELLA. On the contrary, the
size of the field of view of the NOT is about 7 arcmin$^2$. Therefore,
we only observed \mbox{TYC 320-426-1} and fainter reference stars
simultaneously to \hp.

\subsection{Transit data reduction and preparation}

For details on the data reduction, we refer the reader to the
description of DIP$^2$OL \citep{vonEssen2018}. In brief, all the
science frames were bias and flat field calibrated using the IRAF task
{\it ccdproc}. Owed to the large availability of calibration frames,
we corrected the science frames of a given observing night with the
calibrations taken during that particular night only. To produce
photometric light curves we used the IRAF task {\it apphot}. We
measured fluxes inside apertures centered on the target and all
available reference stars within the field of view of the
telescopes. Their sizes were set as a fraction of the full-width at
half-maximum (FWHM) computed and averaged per night. In particular,
the apertures were set between 0.7 and 5 times the FWHM, divided into
10 equally spaced steps. For each one of the apertures we chose three
different sky rings, being their widths 1, 2, and 3 times the
FWHM. The inner radii of the sky ring was fixed to 5 times the FWHM.

Next, we produced differential light curves for the target and each
combination of reference stars by dividing the flux of the target by
the average of a given combination of fluxes from the reference stars.
For each one of the light curves we computed the standard deviation of
the residuals, that were obtained dividing the differential light
curve by a high-degree, time-dependent polynomial that was fitted to
the whole data, including in-transit points, minimizing the sum of the
squares of the residuals. In particular, the degree of the polynomial
was chosen to account for the shallow transit depth and most of the
systematic noise. After visually inspecting all the light curves, it
was chosen to be a septic degree and was fixed along all the
nights. The final combination of reference stars and aperture was
chosen by minimizing the standard deviation of the residual light
curves. This process was performed per transit, individually. With the
time set in Julian dates and the differential fluxes defined, we
finally changed the magnitude of the error bars provided by the {\it
  apphot} task so that their averaged magnitude was coincident with
the standard deviation of the data. Besides time, flux and errors, we
also computed (x,y) centroid positions, integrated flat counts in the
selected aperture around those (x,y) positions, the seeing, the
airmass, and the background counts inside the selected sky ring, all
of them as a function of time and for all the stars involved in the
selected differential light curve. 

\subsection{Data reduction of the long-term monitoring of \hp}

Stellar activity, and particularly stellar spots rotating with the
star, can mimic transit timing variations (see
Section~\ref{sec:stellarvar}). As part of the VAriability MOnitoring
of exoplanet host Stars (VAMOS) project, we carried out a photometric
follow-up of the host star, \hp, along three years. For this end we
observed with the wide-field imager WIFSIP mounted at the 1.2 meter
twin-telescope STELLA, located in the Canary Islands, Spain
\citep{Strassmeier2004,Weber2012}. Table~\ref{tab:STELLA} shows the
main characteristics of the observations. The data were reduced using
standard routines of ESO-MIDAS. On average, around five stars were
used to construct the differential light curve. For details on the
data reduction steps, we refer the reader to \cite{Mallonn2015} and
\cite{Mallonn2016}.

\begin{table}[ht!]
  \caption{\label{tab:STELLA} Main characteristics of the observations
    performed with STELLA. From left to right: time range, $\Delta$t
    in dd.mm.yyyy, photometric bands, PB, exposure time in seconds,
    t$_{exp}$, minimum and maximum exposures taken per filter and per
    night N$_{min/max}$, total number of nightly averaged data points,
    N$_{av}$, and standard deviation of the light curves per observing
    season in ppt, $\sigma_{phot}$.}

  \centering
  \begin{tabular}{c c c c c c}
    \hline \hline
    $\Delta$t               &    PB    &     t$_{exp}$     &   N$_{min/max}$    &   N$_{av}$   &   $\sigma_{phot}$ \\
    \hline
    21.03.2012 - 16.08.2012 &  V       &   30      &   3/36           &   51   &  2.1 \\
    21.03.2012 - 16.08.2012 &  I       &   30      &   3/36           &   45   &  2.1 \\
    09.04.2016 - 29.07.2016 &  B       &   40      &   3/4            &   48   &  2.3 \\
    09.04.2016 - 29.07.2016 &  V       &   30      &   3/4            &   48   &  2.2 \\
    25.02.2017 - 05.08.2017 &  B       &   40      &   3/4            &   67   &  2.3 \\
    25.02.2017 - 05.08.2017 &  V       &   30      &   3/4            &   62   &  2.2 \\
    \hline
  \end{tabular}
  \tablefoot{dd.mm.yyyy corresponds to the day (dd), month (mm)
    and year (yyyy) in which the observations were performed.}
\end{table}

\section{Model parameters and transit analysis}
\label{Sec:Analysis}

\subsection{Choice of detrending model and correlated noise treatment}
\label{sec:detrend}

To clean the data from systematics we constructed an initial
detrending model, DM, with their terms represented by a linear
combination of seeing, S, airmass, $\chi$, (x,y) centroid positions of
all the stars involved in the creation of the differential light
curves, and their respective integrated flat counts and background
counts around these positions, henceforth, the detrending components:

\begin{equation}
  \begin{split}
  DM(t) = a_0 + a_1\cdot S + a_2 \cdot \chi + \,\\
  \sum_1^{N+1} c_i\cdot BG_i + d_i \cdot FC_i + e_i \cdot X_i + f_i \cdot Y_i\,
  \end{split}
\end{equation}

\noindent Here, N is the total number of reference stars. \mbox{N+1}
includes the components of the target star, too. $X_i$ and $Y_i$
correspond to the (x,y) centroid positions, FC$_i$ to the integrated
flat counts, and BG$_i$ to the background counts. The coefficients of
the detrending model are $a_0, a_1, a_2$, and $d_i, e_i$ and $f_i$,
with \mbox{i = 1, N+1}. Besides this linear combination, we also
considered time-dependent polynomials with degrees ranging from one to
three. The detrending strategy is fully described in
\cite{vonEssen2018}. In an iterative process we defined a joint model
conformed by the transit model (see Section~\ref{sec:TFit} for
details) times the detrending model. During each iteration, where we
evaluated the different time-dependent polynomials and the different
combinations of detrending components, we computed reduced $\chi^2$,
the Bayesian Information Criterion \citep[see e.g.,][for details on
  its use]{vonEssen2017} and the weighted standard deviation between
the transit photometry and the combined model. In all these
statistical indicators, the number of fitting parameters are taken
into consideration. Thus, we took care that the data were not being
over-fitted by an unnecessary large detrending model. After averaging
the three statistical indicators, we chose as final detrending model
(and thus, the final number of detrending components or the final
degree for the time-dependent polynomial) the one that minimized this
averaged statistic. Usually the airmass, the (x,y) centroid positions
and the integrated flat counts of all the stars involved in the
construction of the differential light curves were part of the chosen
detrending model. For the Nordic Optical Telescope data, that counts
with an excellent tracking system, a second order, time-dependent
polynomial was usually sufficient to account for instrumental
systematics. Once the detrending model was defined, we treated
correlated noise as specified in \cite{vonEssen2013}. In brief,
following \cite{Carter2009} we produced residual light curves,
subtracting the primary transit model from the data
\citep{MandelAgol2002}. As orbital parameters we used bibliographic
values, listed in this work in the first column of
Table~\ref{tab:tparams}. We then divided each light curve into M bins
of equal duration, each bin holding an averaged number of N data
points. An averaged N accounts for non-equally spaced data points. If
the data is free of correlated noise, then the noise within the the
residual light curves should follow the expectation of independent
random numbers $\hat{\sigma_N}$,

\begin{equation}
  \hat{\sigma_N} = \sigma_1 N^{-1/2}[M/(M-1)]^{1/2}\ .
\end{equation}

\noindent Here, $\sigma_1$ is the sample variance of the unbinned data
and $\sigma_N$ is the sample variance (or RMS) of the binned data:

\begin{equation}
  \sigma_N = \sqrt{\frac{1}{M}\sum_{i = 1}^{M}(<\hat{\mu_i}> - \hat{\mu_i})^2}\ ,
\end{equation}

\noindent where the mean value of the residuals per bin is given by
$\hat{\mu_i}$, and $<\hat{\mu_i}>$ is the mean value of the
means. When correlated noise can not be neglected, each $\sigma_N$
differs by some factor $\beta_N$ from their expectation
$\hat{\sigma_N}$. We computed $\beta$ averaging the $\beta_N$'s
obtained over time bins close to the duration of \hpb's transit
ingress (or, equivalently, egress), which is estimated to be $\sim$15
minutes. To estimate $\beta$, we used as time-bin 0.8, 0.9, 1, 1.1,
and 1.2 times the duration of ingress. Finally, we enlarged the
magnitude of the error bars by $\beta$ \citep{Pont2006}. On average,
CASLEO data presented slightly more correlated noise ($\beta\sim$2.1)
compared to the NOT ($\beta\sim$1.5). This is explained by guiding of
the stars (see column eight of Table~\label{tab:obspars} for a
measurement of the amplitude of variability of the centroid position
of the stars). While the NOT has a stable guiding system, CASLEO lacks
of guiding and the telescope slightly tracks in the East-West
direction, affecting more the quality of the photometry.

\subsection{Transit fitting}
\label{sec:TFit}

With the ten light curves fully constructed, we converted the time
stamps from Julian dates to Barycentric Julian dates, BJD$_{\rm TDB}$,
using the web tool provided by \cite{Eastman2010}. The first step was
to determine the expectation values of the model parameters of
\hpb. For this end, we carried out a Markov-chain Monte Carlo (MCMC)
global fit, using the
\texttt{PyAstronomy}\footnote{\url{http://www.hs.uni-hamburg.de/DE/Ins/Per/Czesla/}\\ \url{PyA/PyA/index.html}}
packages \citep{Patil2010,Jones2001}. As transit model we used the one
provided by \cite{MandelAgol2002}, along with a quadratic
limb-darkening law. In particular, for the \cite{MandelAgol2002}
transit model the fitting parameters are the inclination in degrees,
$i$, the semi-major axis in stellar radius, $a/R_\mathrm{S}$, the
orbital period in days, $Per$, the planet-to-star radii ratio,
R$_\mathrm{P}$/R$_\mathrm{S}$, the mid-transit time in BJD$_{\rm
  TDB}$, T$_\mathrm{0}$, and the linear and quadratic limb darkening
coefficients, u$_1$ and u$_2$.

The quadratic limb-darkening coefficients were taken from
\cite{Claret2000}, for the Johnson-Cousins $R$ filter and stellar
parameters closely matching the ones of
\hp\ \citep[\mbox{T$_{\mathrm{eff}}$ = 5079 $\pm$ 88 K}, \mbox{[Fe/H]
    = -0.04 $\pm$ 0.08}, and \mbox{log(g) = 4.56 $\pm$
    0.06,}][]{Hartman2011}. Thus, the values for the limb-darkening
coefficients used in this work correspond to the stellar parameters
\mbox{T$_{\mathrm{eff}}$ = 5000 K}, \mbox{[Fe/H] = 0} and \mbox{log(g)
  = 4.5}, and are listed in Table~\ref{tab:tparams}. In this work,
quadratic limb darkening coefficients are considered as fixed
values. Nonetheless, to test if the adopted values for the limb
darkening coefficients have an impact on the measured mid-transit
times, we considered the linear and quadratic limb darkening values as
fitting parameters as well. After carrying out a global fit including
them in the fitting process, our posterior distributions for the
parameters revealed values inconsistent with the spectral
classification of the host star. Although our data present a
photometric precision in the part-per-thousand level, the transit
light curves are not precise enough to realistically fit for the limb
darkening coefficients. Nonetheless, mid-transit times computed
fitting for the limb darkening coefficients do not show a significant
offset with respect to the ones computed fixing them, but slightly
larger error bars.

To fit the data with our combined model, we began our analysis
considering the transit parameters listed in \cite{Hartman2011},
\cite{Stevenson2016} and \cite{Wakeford2017} as starting values. The
adopted values are shown in the second column of
Table~\ref{tab:tparams}. We chose uniform priors for the orbital
period, and Gaussian priors for the semi-major axis, the inclination,
and the planet-to-star radii ratio. To be conservative and avoid
misleading the convergence of the fitting algorithm, as standard
deviation for the Gaussian priors we chose three times the magnitude
of the errors determined by the previously mentioned authors. The
mid-transit time determined by \cite{Stevenson2016} has already five
decimals of precision, which translates into an accuracy of $\sim$2
seconds. Since this precision is significantly below our data cadence,
throughout this work the reference mid-transit time is consider as
fixed.

\begin{figure}[ht!]
  \centering
  \includegraphics[width=.5\textwidth]{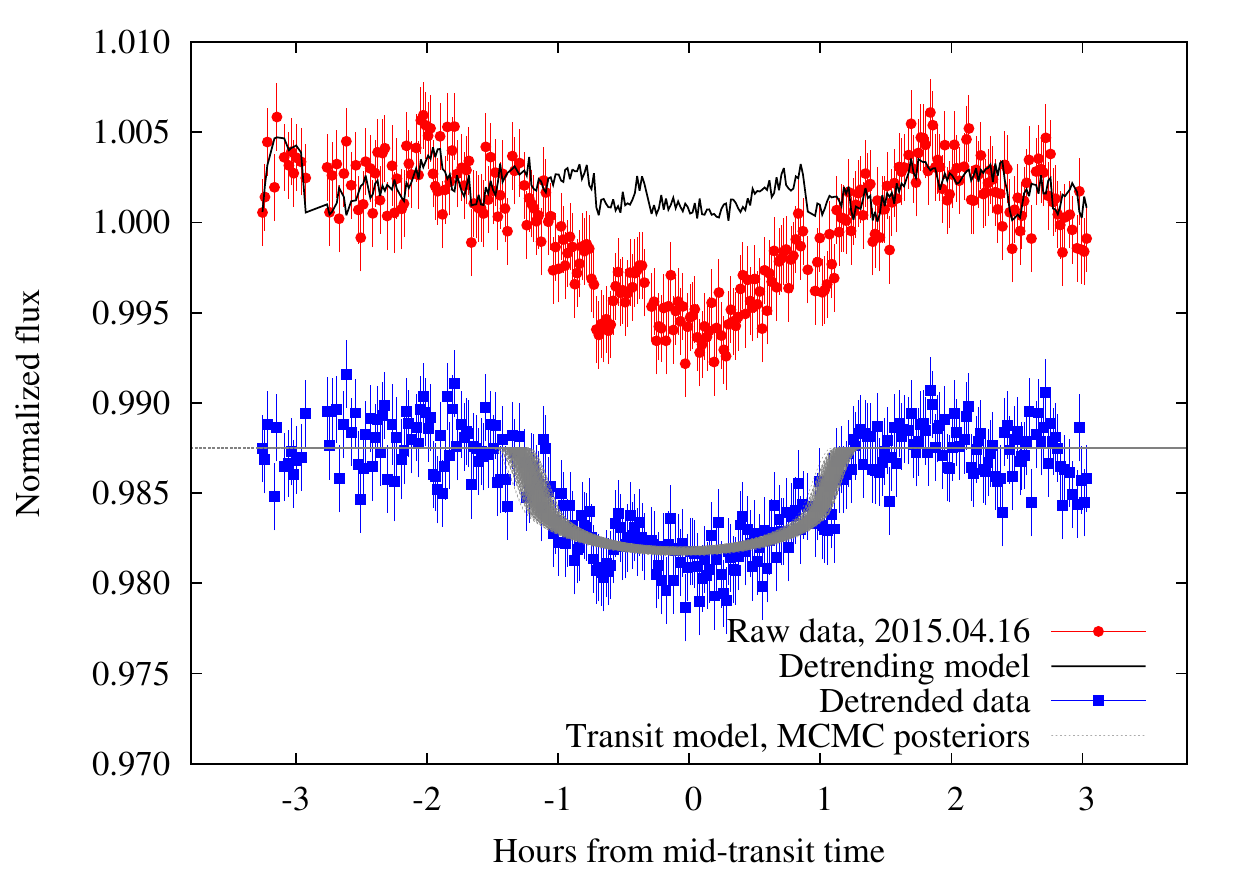}
  \caption{\label{fig:detrendingsteps} Detrending process. The figure
    shows the evolution of the transit photometry during the fitting
    procedure. Raw data are plotted in red circles. The detrending
    model is plotted in black continuous line. Artificially shifted,
    the detrended data are plotted in blue squares, along with primary
    transit models which transit parameters are determined by the
    posterior distributions of the MCMC chains.}
\end{figure}

Once the priors were set, we iterated the fitting process \mbox{1.5
  $\times$10$^5$} times. At each MCMC step not only the transit
parameters are changed, but also the detrending coefficients are
computed with the previously mentioned inversion techniques. Both
transit parameters and detrending coefficients are saved at each MCMC
step. When the total number of iterations were reached, we burned the
initial 50\ 000 samples. From the mean values and standard deviations
of the samples we computed the expectation values of the model
parameters and their corresponding errors. The same process was done
to determine the best-fit detrending coefficients. Throughout this
work we provide the errors of the transit parameters at 1-$\sigma$
level. As a consistency check we analyzed the posterior distributions
to ensure convergence of the chains. To this end, we divided the
chains in three equally large fractions, we computed the model
parameters for each one of the sub-chains, and we checked that they
were consistent within 1-$\sigma$ errors. Our best-fit values are
listed in the third column of Table~\ref{tab:tparams}. To avoid visual
contamination, not all the values for the transit parameters derived
by other authors are displayed on
Table~\ref{tab:tparams}. Nonetheless, the transit parameters presented
in this work are in good agreement with all the bibliographic
values. In particular, the value for the planet-to-star radii ratio
follows the trend observed by \cite{Wakeford2017}, and not the value
reported by \cite{Hartman2011}. Figure~\ref{fig:detrendingsteps} shows
one transit light curve and both model components isolated.

\begin{table}[ht!]
  \caption{\label{tab:tparams} Bibliographic and derived transit
    parameters of \hpb. H11 corresponds to \cite{Hartman2011}, and S16
    to \cite{Stevenson2016}. }
  \centering
  \scalebox{0.95}{
  \begin{tabular}{l c c}
    \hline \hline
    Parameter     & Bibliographic values                       &   This work  \\
    \hline
    Per (days)    & 4.2345023 $\pm$ 7$\times$10$^{-7}$ (S16)    &  4.23450236 $\pm$ 3$\times$10$^{-8}$ \\  
    a/R$_S$        & 11.8 $\pm$ 0.6 (S16)                      &  12.05 $\pm$ 0.13 \\
    i ($^{\circ}$)  & 87.3 $\pm$ 0.4 (S16)                      &  87.31 $\pm$ 0.09 \\
    R$_P$/R$_S$    & 0.0737 $\pm$ 0.0012 (H11)                 &  0.07010 $\pm$ 0.00016 \\
    T$_\mathrm{0}$*  & 5304.65218 $\pm$ 2.5$\times$10$^{-5}$ (S16)           & Adopted from S16 \\
    u$_1$          & -                                        & 0.5140 (fixed) \\
    u$_2$          & -                                        & 0.2180 (fixed) \\ 
    \hline
  \end{tabular}
  }
  \tablefoot{T$_\mathrm{0}$* corresponds to BJD$_{\rm TDB}$ - 2450000.}
\end{table}

\begin{figure*}[ht!]
  \centering
  \includegraphics[width=.85\textwidth]{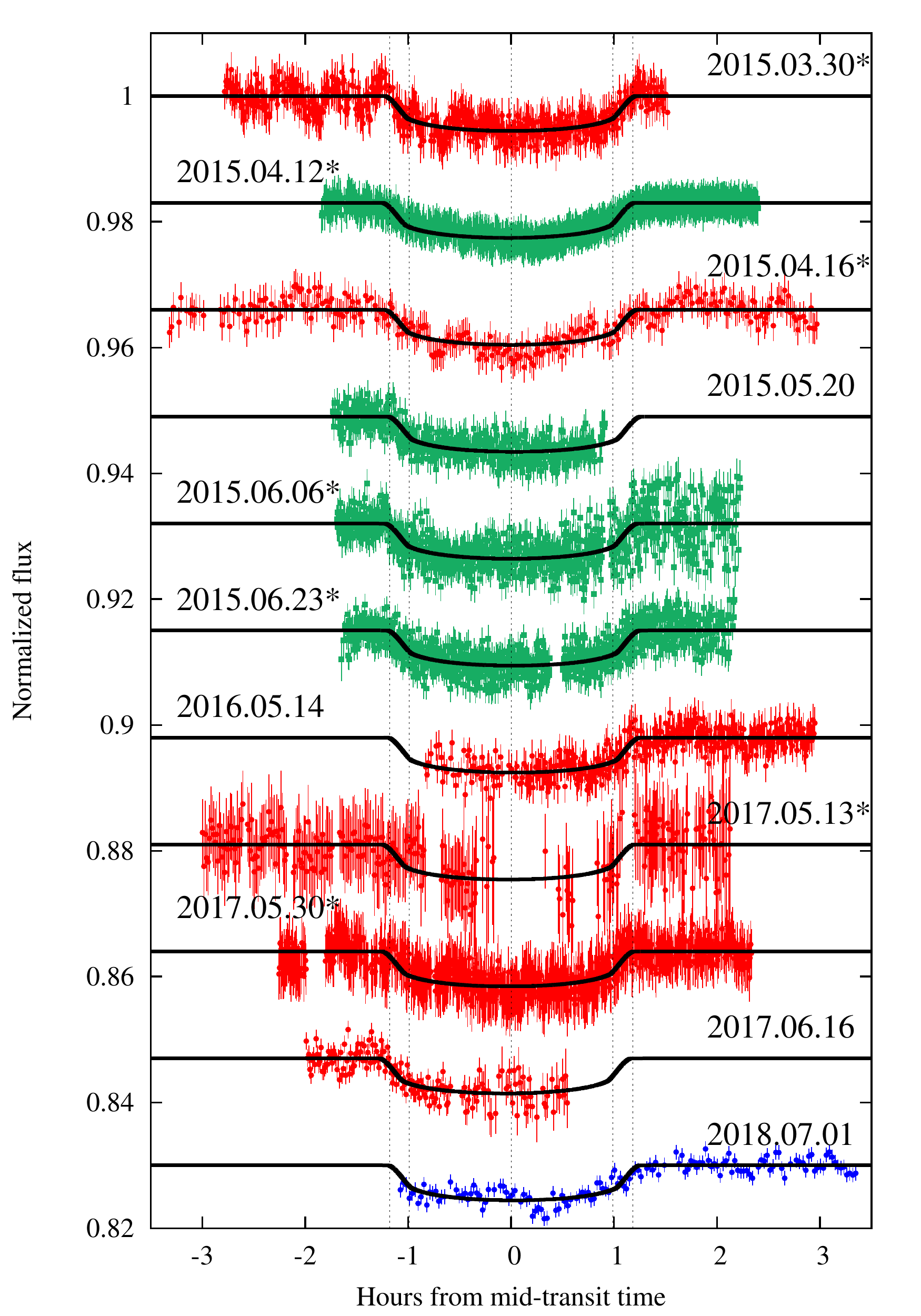}
  \caption{\label{fig:transits} The eleven transit light curves
    collected in this work. The transits observed with the Argentinian
    2.15 meter telescope, CASLEO, are plotted in red circles. The ones
    taken with the Scandinavian 2.5 meter, NOT, are shown in green
    squares. The transit observed with STELLA is shown in blue
    circles. Calender dates are displayed, and asterisk, *, indicate
    the transits that are complete.  Vertical dashed lines indicate
    the four contact times to guide the readers eyes. Transits have
    been folded using our best-fit orbital period. In
      consequence, TTVs can be appreciated by comparing ingress or
      egress to the vertical dashed lines.}
\end{figure*}

After producing a global fit to the complete data set (see
Section~\ref{sec:TFit}), to characterize possible transit timing
variations in the system we proceeded to fit the transits
individually. Figure~\ref{fig:transits} shows the eleven transits,
seven of them correspond to the JS telescope (red circles), one to
STELLA (blue circles) and the remaining four to the NOT (green
squares). To provide TTVs as realistic as possible, for this analysis
we fitted each individual transit. However, for a better assessment of
the TTVs, complete and incomplete transits are clearly
differentiated.

To compute the mid-transit times of the individual light curves,
rather than considering our best-fit model parameters as fixed, we
used our global best-fit values and errors for the mean and standard
deviation respectively for the Gaussian priors. For each individual
mid-transit time we chose uniform priors. Since the determination of
the individual mid-transit times requires the analysis of individual
light curves, the orbital period was considered as fixed to the value
derived from the global fit. This will properly propagate the
uncertainties of all the transit parameters into the determination of
the mid-transit times and, thus, will provide more realistic
uncertainties. Equivalently to the global fit, we iterated and burned
the same number of chains per light curve, and we computed the
individual mid-transit times in the same fashion as specified in
Section~\ref{sec:TFit}. For each light curve we also computed the
transit parameters and we checked that they were consistent with the
global ones, always at 1$\sigma$ level. The first section of
Table~\ref{tab:midtimes} shows the individual mid-transit times from
the bibliography and their respective uncertainties, while the second
section of the same table shows the mid-transit times and the
uncertainties derived in this work.

\begin{table}[ht!]
  \caption{\label{tab:midtimes} From left to right: calender date,
    epoch, mid-transit time in BJD$_{\rm TDB}$ - 2450000 along with
    1$\sigma$ uncertainties, and derived O-C values in minutes without
    uncertainties, to facilitate the comparison with
    Figure~\ref{fig:TTVs}. The first section of the table shows
    bibliographic values taken from \cite{Hartman2011} (H11),
    \cite{Stevenson2016} (S16), and \cite{Wakeford2017} (W17). The
    second section are the values derived in this work. The dates with
    an * correspond to mid-transit times derived from complete
    transits.}  \centering \scalebox{0.85}{
  \begin{tabular}{l c c c c}
    \hline
    \hline
    Date       & Epoch  &  Mid-transit time        & Uncertainty & O-C\\
    yyyy.mm.dd &        &  BJD$_{\rm TDB}$ -2450000  &  (days)     & (minutes) \\             
    \multicolumn{3}{l}{Bibliographic values} \\
    \hline 
    2009.01.28 (H11) & -105 & 4860.02786  & 0.00147    &  -3.3 \\
    2010.04.18 (S16*) &    0 & 5304.65218  & 0.000025  &  -1.4 \\ 
    2010.05.25 (H11) &    9 & 5342.76262  & 0.00041    &  -1.2 \\
    2013.04.23 (S16) &  260 & 6405.6237   & 0.0009     &   1.3 \\
    2013.09.09 (S16) &  293 & 6545.3622   & 0.0003     &   1.2 \\
    2015.04.16 (S16) &  431 & 7129.72248  & 0.00017    &  -0.3 \\ 
    2016.01.25 (W17) &  498 & 7413.432836 & 0.000172   &  -2.2 \\
    2016.03.12 (W17) &  509 & 7460.013266 & 0.000016   &  -0.9 \\
    2016.05.02 (W17) &  521 & 7510.827100 & 0.000016   &  -1.2 \\
    2016.08.16 (W17) &  546 & 7616.690103 & 0.000011   &  -0.5 \\
    \hline \hline
    \multicolumn{4}{l}{Our work} \\
   \hline
    2015.03.30* & 427 & 7112.78503 &  0.00058          & 0.5  \\
    2015.04.12* & 430 & 7125.48930 &  0.00063          & 1.6  \\
    2015.04.16* & 431 & 7129.72283 &  0.00063          & 0.2  \\
    2015.05.20  & 439 & 7163.59738 &  0.00040          & -1.9  \\
    2015.06.06* & 443 & 7180.53670 &  0.00057          & -0.03  \\
    2015.06.23* & 447 & 7197.47376 &  0.00046          & -1.4  \\
    2016.05.14  & 524 & 7523.53041 &  0.00072          & -1.5  \\
    2017.05.13* & 610 & 7887.69984 &  0.00089          & 1.7  \\
    2017.05.30* & 614 & 7904.63796 &  0.00066          & 1.9  \\
    2017.06.16  & 618 & 7921.57698 &  0.00078          & 3.4  \\
    2018.07.01  & 690 & 8226.45772 &  0.00093          & -1.6  \\
    \hline
  \end{tabular}
  } \tablefoot{ The two transit times from H11 are not individually
    measured transit times, but rather two reference times which are
    used in place of the period and reference epoch when
    simultaneously fitting all of the data, and assuming a constant
    linear ephemeris. Since both epochs differ, this will produce a
    spurious offset in the O-C diagram (see epoch zero in
    Figure~\ref{fig:TTVs}) unrelated to TTVs.}
\end{table}

\section{Analysis and results}
\label{Sec:TSQ}

\subsection{Transit timing variations}

\begin{figure}[ht!]
  \centering
  \includegraphics[width=.5\textwidth]{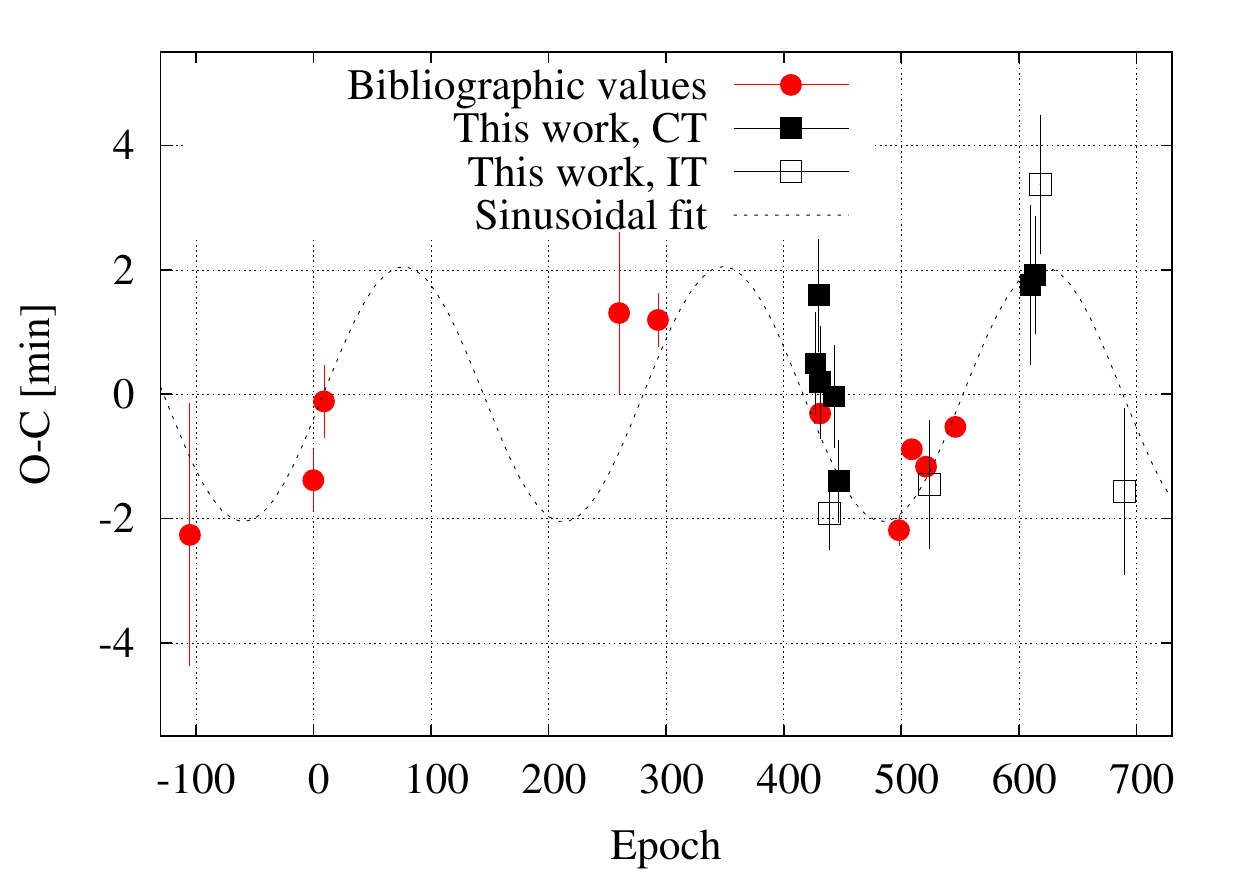}
  \caption{\label{fig:TTVs} Observed minus calculated (O-C)
    mid-transit times for \hpb\ in minutes. Red circular points show
    the values obtained from \cite{Hartman2011}, \cite{Stevenson2016},
    and \cite{Wakeford2017}. Black filled squares correspond to the
    values derived in this work considering complete transits (CT),
    while empty squared are those derived from incomplete transits
    (IT). The O-C diagram was computed using the best-fit orbital
    period derived in this work, specified in
    Table~\ref{tab:tparams}. The dashed black line corresponds to our
    best-fit sinusoidal variability.}
\end{figure}

We computed the TTVs in \hpb's system by subtracting to the observed
mid-transit times the ones obtained from a constant (best-fit) orbital
period. The derived values can be found on Figure~\ref{fig:TTVs} and
the lower part of Table~\ref{tab:midtimes}. Note that in the figure
the zeroth epoch does not lie over the abscissas axis. This is due to
the improvement of its value between \cite{Hartman2011} and
\cite{Wakeford2017}.

In the absence of any timing variations we expect no significant
deviations of the derived O-C values from zero. To test the null
hypothesis of \mbox{O-C = 0}, we used a $\chi^2_{red}$ test. For 20
degrees of freedom the value raises up to \mbox{$\chi^2_{red}$ = 177},
with a p-value, \mbox{P$<$1$\times$10$^{-5}$}. For this, the
significance level is chosen to be 1\%. This simple statistical
analysis, $\chi^2_{red}$, and its corresponding p-value indicate that
the mid-transit times of \hpb\ are inconsistent with a constant
period. A visual inspection of \hpb's O-C diagram suggests that the
TTVs are strongly dependent on the last three data points. To
test these assumptions we excluded the three last epochs and
re-computed our statistics. A \mbox{$\chi^2_{red}$ = 102},
\mbox{P$<$1$\times$10$^{-5}$}.

For completeness, assuming TTVs with a sinusoidal variability we
fitted to our timing residuals a simple sinusoidal function with the
following expression:

\begin{equation}
\mathrm{TTVs}(E) = A_{\mathrm{TTVs}} \sin[2\pi(E/Per_{\mathrm{TTVs}} - \phi_{\mathrm{TTVs}})]\,,
\end{equation}

\noindent where $E$ corresponds to the epoch, $A_{\mathrm{TTVs}}$
corresponds to the semi-amplitude of the timing residuals,
$Per_{\mathrm{TTVs}}$ to the periodicity, and $\phi_{\mathrm{TTVs}}$
to the phase. From our analysis, we obtain
\mbox{$A_{\mathrm{TTVs}}\sim$2.1 minutes}, while the best-fit
periodicity would correspond to around 270 epochs. This is equivalent
to almost 1100 days. Comparing the semi-amplitude with the average
error bars for the timings we find
$A_{\mathrm{TTVs}}\sim$3$\times\epsilon$, being $\epsilon$ the
averaged value of the timing errors. Our best-fit sinusoidal function
is plotted in Figure~\ref{fig:TTVs} with a black dashed line. Mainly
due to possible aliasing effects, and due to the complications that
single-transiting planets with TTVs convey, we believe any analysis on
the TTVs reported in this work are subject to strong
speculations. More spectroscopic and photometric data are required to
characterize the body causing these TTVs.

\subsection{TTVs induced by stellar activity}
\label{sec:stellarvar}

Due to the high photometric quality provided by space-based
observations such as CoRoT \citep{Auvergne2009} and Kepler, stellar
magnetic activity and its impact on transit light curves have been
studied in great detail. Dark spots and bright feculae on the stellar
photosphere move as the star rotates, producing a time-dependent
variation of the stellar flux. The photometric precision we can
achieve nowadays is so high, that the small imprints of occulted spots
on transit data have been used to characterize stellar surface
brightness profiles and spot migration and evolution \citep[see
  e.g.,][]{Carter2011,Sanchis-Ojeda2011a,Sanchis-Ojeda2011b,Huber2010}.
When transit fitting is performed, an incorrect treatment of these
``bumps'' can have a significant impact on the computation of
planetary and stellar parameters \citep[see
  e.g.,][]{Czesla2009,Lanza2011}.

Occulted and unocculted spots can asymmetrically modify the shape of
the transit light curves, and thus affect the true locations of the
mid-transit times. Actually, the deformations that stellar activity
produces on primary transits have been studied in detail and
pinpointed in some cases as a misleading identification of TTVs
\citep[see e.g.,][]{Rabus2009}. For example, \cite{Maciejewski2011}
found TTVs in the WASP-10 system with an amplitude of few
minutes. They attributed the variability to the gravitational
interaction between the transiting planet and a second body of a tenth
of a Jupiter mass close to a 5:3 mean motion resonance. Although
\cite{Barros2013} did not find a linear ephemeris as a statistically
good fit to the mid-transit times of WASP-10b, they showed that the
observed variability could be produced by, for example, spot
occultation features.

\begin{figure}[ht!]
  \centering
  \includegraphics[width=.5\textwidth]{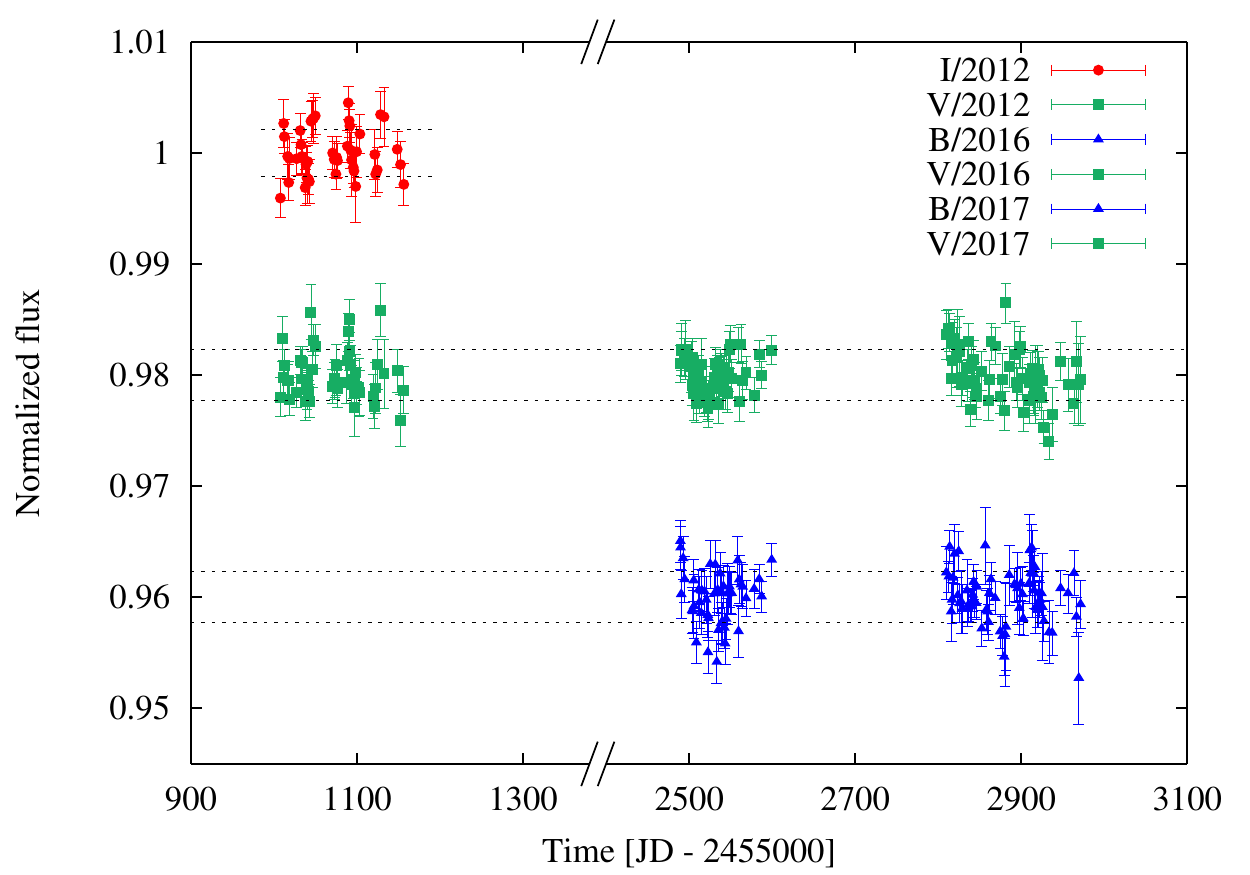}
  \caption{\label{fig:STELLA} Photometric follow-up of \hp\ using the
    STELLA photometer. Red circles correspond to the $I$ band, green
    squares to the $V$ band, and blue triangles to the $B$ band. The
    fluxes have been artificially shifted, and the time axis has been
    compressed for a better visualization. Dashed black lines show the
    averaged standard deviation.}
\end{figure}

To quantify stellar activity in our system and support (or overrule)
our TTV detection, we carried out a photometric follow-up of
\hp\ along three years \citep[see e.g.,][Figure~\ref{fig:STELLA} and
  our Table~\ref{tab:STELLA}]{Mallonn2015,Mallonn2016}. The data,
taken in the Johnson-Cousins B, V, and I filters, show a maximum
scatter of 2.3 ppt. To search for any significant periodicity
contained in the data we computed Lomb-Scargle periodograms
\citep{Lomb,Scargle,LombScargle}, finding featureless periodograms
within each observing season and within the whole observing run. As a
consequence, the photometric follow-up of the host star seems to show
no evidence of spot modulation. Nonetheless, the characterization
presented in this work is as good as the data are. In other words, if
spot induced modulation should exist, it would be well contained
within the $\sim$2 ppm limit.

\cite{Hartman2011} characterized the chromospheric activity of the
star derived from the flux in the cores of the Ca II H and K lines,
the so-called S index, \mbox{S$_{HK}$ = 0.182 $\pm$ 0.004}. Comparing
this value to the relation between the S index and the stellar
brightness variation found by \cite{Karoff2016} (their Figure 5),
observational evidence should place the photometric variability of
\hp\ to be around 1 ppt. Assuming that spot modulation around and
below this limit actually exists, and that it might have an impact on
the TTVs, \cite{Ioannidis2016} found that the maximum amplitude of
TTVs generated by spot crossing events strongly depends, among others,
by the transit duration. In the case of \hpb\ ($\sim$143 minutes),
TTVs caused by spots should have a maximum amplitude of $\sim$1
minute. This is significantly below the TTV amplitude detected in this
work. Therefore, the data presented here seem to support the scenario
of TTVs caused by gravitational interactions instead of modulated by
the activity of the star, if any.

\subsection{TTVs caused by systematic effects}

Besides stellar activity, TTVs can be caused by systematic effects not
properly accounted for. This has a special relevance when
low-amplitude TTVs are being reviewed. A manifestation of this effect
is the derivation of underestimated error bars for the mid-transit
times. While in this work we have taken special attention to the
computation of uncertainties, we can only trust that the values
reported by other authors have had a similar treatment. Nonetheless,
as a consistency check we enlarged the error bars of all the observed
mid-transit times by a factor and re-computed $\chi^2_{red}$ for each
one of the artificially enlarged O-C diagrams. A factor of 13 was
required for $\chi^2_{red}$ to be consistent with 1, equivalently,
with no TTVs. This marginally large value seems to support our TTV
detection.

\section{Discussion and conclusions}
\label{Sec:DaC}

Since its discovery, some indications of variability in both
spectroscopic and photometric data pinpointed \hpb\ as an interesting
target for additional photometric follow-up. For the last three years
our group has been collecting primary transit data of
part-per-thousand photometric precision, necessary to detect the
shallow transits that the planet produces while crossing its host star
each $\sim$4.23 days. In this work we have reduced all the new data in
an homogeneous way, we have updated and improved the transit
parameters and ephemeris, and we have detected a significant
variability in the timing residuals. Furthermore, we have followed the
host star along several years to characterize its activity and, if
observed, its impact in the mid-transit times. However, light curves
taken in three different bands along three years revealed no spot
modulation within the precision limit of the data, devaluing the
scenario of spot-induced TTVs. Due to the complexity of the analysis
of single-transiting systems presenting TTVs, it is hard to drop any
conclusions on the characteristics of the perturbing
body. Nonetheless, we understand these results as relevant to motivate
future photometric and spectroscopic follow-up campaigns, that will
contribute to disclose the origin of the observed variability.

\begin{acknowledgements}

Visiting Astronomer, Complejo Astron\'omico El Leoncito operated under
agreement between the Consejo Nacional de Investigaciones
Cient\'ificas y T\'ecnicas de la Rep\'ublica Argentina and the
National Universities of La Plata, C\'ordoba and San Juan. CvE
acknowledges funding for the Stellar Astrophysics Centre, provided by
The Danish National Research Foundation (Grant DNRF106), and support
from the European Social Fund via the Lithuanian Science Council grant
No. 09.3.3-LMT-K-712-01-0103. Based on observations made with the
Nordic Optical Telescope, operated by the Nordic Optical Telescope
Scientific Association at the Observatorio del Roque de los Muchachos,
La Palma, Spain, of the Instituto de Astrofisica de Canarias. We
acknowledge support from the Research Council of Norway's grant 188910
to finance service observing at the NOT. SW acknowledges support for
International Team 265 (``Magnetic Activity of M-type Dwarf Stars and
the Influence on Habitable Extra-solar Planets'') funded by the
International Space Science Institute (ISSI) in Bern, Switzerland. VP
wishes to acknowledge the funding that was provided in part by a
J. William Fulbright Grant and the sponsorship of
Georg-August-Universitat in Gottingen, Germany. We thank the referee
for the comments that contributed to improve this work. This work made
use of
PyAstronomy\footnote{\url{https://github.com/sczesla/PyAstronomy}}.

\end{acknowledgements}

\bibliographystyle{aa}
\bibliography{vonEssenC}
\end{document}